\documentclass[reprint,amsmath,amssymb,aps,prx,balancelastpage,floatfix,
bibnotes]{revtex4-2}
\UseRawInputEncoding
\usepackage{graphicx}
\usepackage{dcolumn}
\usepackage{bm}
\usepackage{mathtools}
\usepackage{hyperref}
\usepackage{hycolor}
\hypersetup{colorlinks=true,citecolor=[rgb]{0,0.15,0.60}, 
urlcolor=[rgb]{0,0.15,0.60}, linkcolor=[rgb]{0,0.20,0.80}, final = true}
\usepackage{enumitem}
\usepackage[rightcaption]{sidecap}
\usepackage{physics}
\usepackage{relsize}
\usepackage{siunitx}
\DeclareSIUnit{\Poise}{P}
\DeclareSIUnit{\frames}{frames}
\newcommand*{\balancecolsandclearpage}{%
  \close@column@grid
  \clearpage
  \twocolumngrid
}
\setcitestyle{round,super}

\def\ith{$i^{\mathrm{th}}$~}

\def\jth{$j^{\mathrm{th}}$~}

\def\half{{\textstyle\frac{1}{2}}}
\def\halfsqthree{{\textstyle\frac{\sqrt{3}}{2}}}
\def\twothirds{{\textstyle\frac{2}{3}}}
\def\sixth{{\textstyle\frac{1}{6}}}

\def\eg{{\it{e.g.,~}}} 
\def\ie{{\it{i.e.,~}}}

\def\brb{\bar{\rb}}
\def\ab{\mathbf{a}}
\def\bb{\mathbf{b}}

\def\fb{\mathbf{f}}
\def\Fb{\mathbf{F}}
\def\kb{\mathbf{k}}
\def\rb{\mathbf{r}}
\def\ub{\mathbf{u}}
\def\Db{\mathbf{D}}
\def\Omegab{\boldsymbol{\Omega}}
\def\sigmab{\boldsymbol{\sigma}}
\def\ubp{\ub_\mathrm{p}}

\def\dkb{\mathbf{dk}}
\def\br{\bar{r}}
\def\delrb{\mathbf{\delta r}}

\def\psik{\boldsymbol{\psi}_\kb}
\def\Psik{\Psi_\kb}
\def\kD{\kb_\Db}
\def\fbij{\fb^{ij}}
\def\fbji{\fb^{ji}}
\def\xij{x_{ij}}
\def\yij{y_{ij}}
\def\rij{r_{ij}}
\def\rbij{\rb_{ij}}

\def\brij{\br_{ij}}

\def\brbj{\brb_j}
\def\brj{\br_j}

\def\HH{\mathbf{H}}
\def\Hk{\mathcal{H}_\kb}
\def\FF{\mathcal{F}}
\def\Hij{\HH^{ij}}

\def\tetij{\theta_{ij}}
\def\tetji{\theta_{ji}}
\def\tetj{\theta_{j}}
\def\Sk{S(\kb)}

\def\sigx{\sigma_{x}} 
\def\sigy{\sigma_{y}}
\def\sigz{\sigma_{z}}

\def\Omx{\Omega_{x}}
\def\Omy{\Omega_{y}}
\def\omk{\omega_\kb}
\def\r{\right}
\def\l{\left}
\def\rang{\r\rangle}
\def\lang{\l\langle}

\def\Rey{\mathrm{Re}}
\def\Pec{\mathrm{Pe}}

\def\T{\intercal}
\def\sb#1{\textbf{\textsf{#1}}}

\def\PSD{\mathrm{PSD}}

\usepackage{xcolor}
\definecolor{YKB}{rgb}{0.1,0.31,0.80}
\definecolor{saff}{rgb}{0.98,0.26,0.14}

\def\movieOneExp{https://tinyurl.com/yckvj7tx}
\def\movieTwoSqMelt{https://tinyurl.com/2p8uaynk}
\def\movieThreeAvalanch{https://tinyurl.com/2prrttwj}
\def\movieFourHexMelt{https://tinyurl.com/vpsh429c}
\def\movieFiveFlatband{https://tinyurl.com/3f55a8cp}

\begin{document}
\title{\larger[1.5] \textsf{Quasiparticles, Flat Bands, and the Melting of Hydrodynamic Matter}}
\author{Imran Saeed}
\affiliation{Center for Soft and Living Matter, Institute for Basic Science (IBS), Ulsan, 44919, Republic of Korea}
\affiliation{Department of Physics, Ulsan National Institute of Science and Technology, Ulsan, 44919, Republic of Korea}
\author{Hyuk Kyu Pak}
\email{hyuk.k.pak@gmail.com}
\affiliation{Center for Soft and Living Matter, Institute for Basic Science (IBS), Ulsan, 44919, Republic of Korea}
\affiliation{Department of Physics, Ulsan National Institute of Science and Technology, Ulsan, 44919, Republic of Korea}
\author{Tsvi Tlusty}
\email{tsvitlusty@gmail.com}
\affiliation{Center for Soft and Living Matter, Institute for Basic Science (IBS), Ulsan, 44919, Republic of Korea}
\affiliation{Department of Physics, Ulsan National Institute of Science and Technology, Ulsan, 44919, Republic of Korea}

\date{October 21, 2021}
\maketitle  
\onecolumngrid
\noindent\textsf{The concept of quasiparticles---long-lived low-energy particle-like excitations---has become a keystone of condensed quantum matter, where it explains a variety of emergent many-body phenomena, such as superfluidity and superconductivity. Here, we use quasiparticles to explain the collective behavior of a classical system of hydrodynamically interacting particles in two dimensions. In the disordered phase of this matter, measurements reveal a sub-population of long-lived particle pairs. Modeling and simulation of the ordered crystalline phase identify the pairs as  quasiparticles, emerging at the Dirac cones of the spectrum. The quasiparticles stimulate supersonic pairing avalanches, bringing about the melting of the crystal. In hexagonal crystals, where the intrinsic threefold symmetry of the hydrodynamic interaction matches that of the crystal, the spectrum forms a flat band dense with ultra-slow, low-frequency phonons whose  collective interactions induce a much sharper melting transition. Altogether, these findings demonstrate the usefulness of concepts from quantum matter theory in understanding many-body physics in classical dissipative settings. }\\
\twocolumngrid

\noindent The idea of quasiparticles was introduced by Landau in 1941,\cite{Landau1941} and ever since has provided insight into emergent collective phenomena in a wide variety of physical settings.\cite{Nozieres1964, Schrieffer1964,Weinberg1963,Hybertsen1985,Woelfle2018} Broadly speaking, quasiparticles are long-lived excitations that behave as weakly interacting particles.\cite{Pines1963,Schrieffer1970} The quasiparticle zoo keeps expanding,\cite{Bradlyn2016,Venema2016,Goebel2021,Rivera2020} with fractional,\cite{Saminadayar1997,Nayak2008} Dirac,\cite{Novoselov2005,Bostwick2007} and Majorana\cite{Wilczek2009,Aguado2017} quasiparticles among the notable new species. Here, we borrow this intuitive notion of quantum matter theory to explain many-body phenomena in a classical 2D system of hydrodynamically interacting particles. 
By combining experiments, simulations, and a theoretical model, we identify long-lived particle pairs as the elementary low-frequency excitations in the system -- its effective quasiparticles that induce crystal melting. At a critical point when the intrinsic threefold symmetry of the hydrodynamic interaction matches that of the hexagonal crystal, the system produces a monkey-saddle van Hove singularity (vHS),~\cite{VanHove1953,Efremov2019,Volovik2017} with a nearly flat band of slow collective excitations that induce a much sharper melting transition. Flat bands and vHSs attracted much interest in recent years,~\cite{Volovik2013,Leykam2018,Yuan2019,Rosenzweig2020,Maimaiti2021} as the slowing down of excitations gives rise to strong correlations associated with high-temperature superconductivity~\cite{Kopnin2011,Yao2015,Mondaini2018,Isobe2018} and topological insulators.~\cite{Bergholtz2013,Tang2014} In particular, graphene bilayers exhibit vHSs,~\cite{Li2009,Efremov2019} flat bands, and pairing~\cite{Christos2020} when twisted,~\cite{Bistritzer2011,Andrei2020} put on a SiC substrate,~\cite{Marchenko2018} or buckled.~\cite{Mao2020} The present work proposes accessible hydrodynamic analogs of these phenomena in the highly-dissipative regime, where exceptional topology has been recently demonstrated.~\cite{Tlusty2021} \\

\begin{figure*}[!htbp]
\centering
 \includegraphics[width=0.95\textwidth]
 {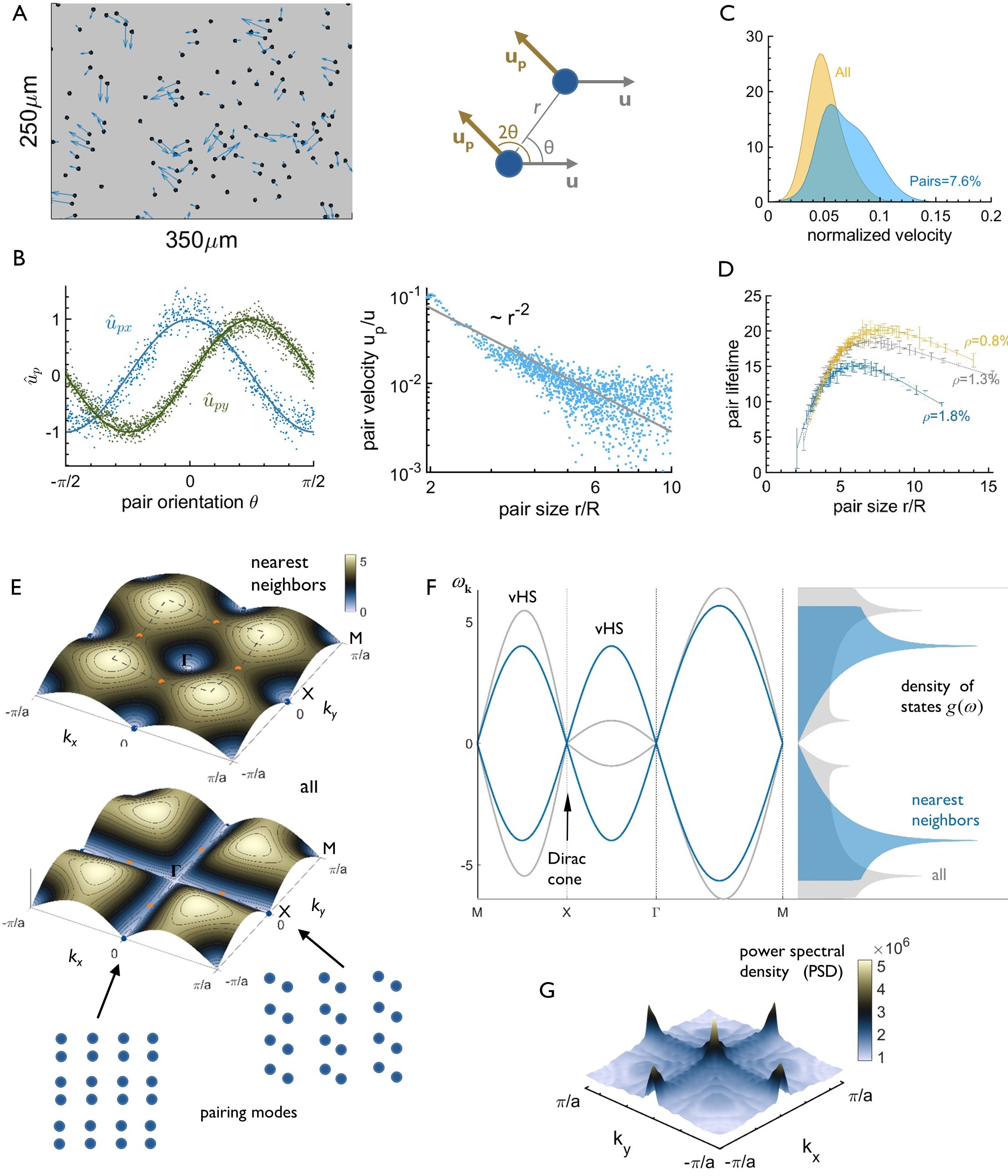}
 \caption{ \sb{Hydrodynamic pairing in disordered and ordered phases.}
\sb{A.} The experimental system with $\ell = \SI{7}{\um}$ particles whose areal density is $\rho=\SI{1.8}{\percent}$ in a channel of height \SI{10}{\um} (left). Arrows denote particle velocity w.r.t. the mean velocity $\ub \simeq \SI{200}{\um/s}$. Notable are pairs of particles moving at similar velocity, as depicted in the schematic (right). 
\sb{B.} The measured pair velocity $\ubp$, showing the direction $\hat{\ub}_{\rm p} \sim (\cos{2 \theta}, \sin{2 \theta})$ (left) and the magnitude $\abs{\ubp} \sim r^{-2}$ (right). Solid lines are the theoretical predictions.
\sb{C.} Distribution of velocity w.r.t. center of mass (in units of $u$) of all particles (gold) and in the pairs (blue, \SI{7.6}{\percent} of all particles).
\sb{D.} Lifetime of pairs (in $R/u$ units) as a function of pair size $r/R$. Bars denote standard error. Between \num{2e5} and \num{7.3e5} pairs were measured in each experiment. 
\sb{E.} Top: The frequency bands $\omk^{+} = -\omk^{-}$ of a square lattice with nearest neighbor interactions, in the first Brillouin zone (BZ), with four Dirac points (blue), and four vHSs inside the BZ (orange). The effective primitive cell shown in dashed black square. 
Bottom: The spectrum with all hydrodynamic interactions, with typical pairing modes depicted.
\sb{F.} The band structure and corresponding density of states $g(\omega)$ (a.u.).
\sb{G.} PSD in a simulation starting from a white-noise perturbation shows strong selection of the pairing modes at the Dirac points. }
\label{fig:1-pairing}
\end{figure*}

\noindent
\sb{Hydrodynamic pairing} \\
 We examine ensembles of polystyrene colloids of diameter $\ell= 2 R = \SIrange{7}{20}{\um}$, confined between the parallel plates of a thin microfluidic channel of height $\SIrange{10}{30}{\um}$. The height is designed to be slightly larger than the particles' diameter, such that their horizontal $(x,y)$ positions cannot overlap, rendering the system effectively two-dimensional, with an areal density of the particles $\rho \simeq \SIrange{0.3}{7}{\percent}$ (Fig.~\ref{fig:1-pairing}A, Methods, and \href{\movieOneExp}{Movie 1}). A steady flow of water drives the particles in the $x$-direction. Slowed down by friction and viscous shear forces at the floor and ceiling of the channel, the particles move at a velocity $u =\SIrange{100}{400}{\um/\s}$ relative to the water, and thus experience a drag force $\gamma u$, with the drag coefficient $\gamma$. To make the particles flow, the momentum loss needs to be compensated by constantly pumping momentum through the pressure gradient along the channel.~\cite{Beatus2006,Beatus2012,Beatus2017} The Reynolds and P{\'e}clet numbers were $\Rey \simeq \numrange{e-4}{e-3}$ and $\Pec \simeq \numrange{e3}{e4}$, allowing us to safely disregard the inertial and thermal forces. 

 As the driven particles are slower than the surrounding water, they perturb the streamlines. In the quasi-2D geometry of our setup, the perturbations are known to induce long-range dipolar interactions.~\cite{Liron1976, Cui2004,Beatus2006,Beatus2007,Baron2008,Beatus2008,Beatus2009,Champagne2011,Beatus2012,Liu2012,Desreumaux2013,Uspal2013,Shani2014,Nagar2014,Shen2016,Tsang2016,Beatus2017} The hydrodynamic force $\fb(\rb)$ exerted on a particle by another particle at a distance $\rb = (r,\theta)$, where $\theta$ is the angle with respect to the flow direction, has a magnitude decaying as the distance squared, $r^{-2}$, and is oriented at twice the angle, $2\theta$ (Fig.~\ref{fig:1-pairing}A, Methods).
This twofold symmetry implies that the dipolar force is \emph{invariant under parity}, $\fb(-\rb) = \fb(\rb)$.~\cite{Beatus2006,Sprott2009} Thus, the hydrodynamic forces that a pair of particles exert on each other are equal, and isolated pairs should therefore be stable. A pair oriented at an angle $\theta$ moves at a velocity $\ubp \sim u (R/r)^2 (\cos{2\theta},\sin{2\theta}$), as verified in the experiment (Fig.~\ref{fig:1-pairing}B). Note that pairs can be stable only thanks to the dissipative nature of the forces. Momentum conserving forces, in contrast, would be anti-symmetric by Newton's third law, $\fb(-\rb) =  -\fb(\rb)$, and thereby destabilize the pairs. From the parity symmetry of the hydrodynamic dipoles originate all physical phenomena described in this article.

Due to the inverse-square decay of the hydrodynamic force, intra-pair forces are typically much stronger than interactions with the surrounding particles, and one would expect to see  weakly-interacting metastable pairs. Analysis of particle trajectories verified this prediction: a significant fraction of the particles, typically about \SIrange{5}{20}{\percent}, traverse in pairs, geometrically defined as couples of particles much closer to each other than to the next-nearest neighbor---by a factor of ${\sim}3.5$, such that their interactions with other particles are at least tenfold weaker (Fig.~\ref{fig:1-pairing}C). The pairs move significantly faster than the whole population (relative to the center of mass). These weakly-interacting couples persist through typical lifetimes ${\sim}\SIrange{10}{20}{\it R/u}$, until they approach other particles (Fig.~\ref{fig:1-pairing}D, \href{\movieOneExp}{Movie 1}).
To exclude the possibility that the system is significantly affected by non-hydrodynamic interactions, such as van der Waals or electrostatic forces, we compared the measurements to simulations of particle ensembles with purely hydrodynamic interaction (and hard-core repulsion, Methods), which exhibited similar velocity and lifetime distributions (Fig.~\ref{fig:S1-comparison}). \\

\noindent\sb{Emergence of quasiparticles in hydrodynamic crystals}\\
\noindent The emergence of pairs observed in the disordered phase hints that these might be elementary particle-like excitations in the system. To explore this possibility, we consider a driven hydrodynamic crystal made of identical particles. A method of generating large hydrodynamic lattices of hundreds to thousands of particles is yet to be developed (though densely-packed~\cite{DelGiudice2021} or spatially-structured~\cite{Pompano2011} microfluidic crystals can be produced by various techniques). Thus, we investigate the ordered crystalline phase using the analytic model and computer simulations tested against the experiment in the disordered phase.

At steady state, the viscous drag force experienced by each particle in the crystal is counterbalanced by the driving force, $\Fb$, and the crystal flows uniformly at a velocity  $\ub= \Fb/\gamma$ (Methods).
The long-range hydrodynamic forces excite collective modes in the lattice. Expanding the dynamical equation in small deviations around the steady-state motion,  
we find that these normal modes are plane waves of wavevector $\kb$, with a polarization $\psik$ and frequency $\omk$, which are the eigenvector and eigenvalue of a Schr{\"o}dinger-like equation, $\Hk  \psik = \omk \psik$.~\cite{Tlusty2021} The hydrodynamic ``Hamiltonian" is
\begin{equation}
       \Hk  =  \Omegab_\kb \cdot \sigmab  ~,
        \label{eq:schrodinger}
\end{equation}
where $\sigmab = (\sigx, \sigy)$ are Pauli's matrices, and $\Omegab_\kb = (\Omx,\Omy)$ are Fourier sums of the hydrodynamic interactions over the steady-state lattice positions $\brbj = (\br_j,\tetj)$,
\begin{equation}
    \Omegab_\kb 
    \equiv 
    \begin{bmatrix}
    \Omx \\ \Omy
    \end{bmatrix} 
    \sim  \sum_{j} 
    \frac{e^{i\kb \cdot \brbj}}{\brj^3}
    \begin{bmatrix}
    \cos{3 \tetj} \\  \sin{3 \tetj}
    \end{bmatrix} ~. 
    \label{eq:Omegas}
\end{equation}
The Hermitian operator $\Hk$ exhibits two purely real eigenfrequency bands,  $\omk =  \pm  \abs{\Omegab_\kb} = \pm(\Omx^2+\Omy^2)^{1/2}$, corresponding to marginally stable \emph{phonon} modes that propagate without any damping (Methods).~\cite{Beatus2006,Beatus2012,Beatus2017} 

Notably, while the dipolar force shows twofold symmetry ($2\theta$ in Fig.~\ref{fig:1-pairing}B), the rotational symmetry of $\Hk$ is \emph{threefold} ($3\theta$ in Eq.~(\ref{eq:Omegas})).
This is because $\Hk$ is the momentum-space ``spring constant", linking the stress and the strain in the hydrodynamic crystal (Methods). Thus, since the force is dipolar,~\cite{Beatus2012} $\fb \sim e^{i 2\theta}/r^2$, the spring constant $\Hk$ is tripolar, $\Hk \sim \sum_\rb{(\grad_\rb{\fb})e^{i\kb\rb} } \sim \sum_\rb{ (e^{i 3\theta}/r^3) e^{i\kb\rb}}$ (Eq.~(\ref{eq:Omegas})), which has interesting implications on the collective modes, as discussed in the following.

It is instructive to consider first a simple case, a square crystal of lattice constant $a$ in which particles interact only with their nearest neighbors, and the resulting frequency bands are $\omk \sim  \pm \sqrt{ 2 - \cos{2 k_x a} -\cos{2 k_y a} }$ (Fig.~\ref{fig:1-pairing}E,F). Two distinctive features of the spectrum are: (i) four Dirac points 
(X points, $\kb = \kD$), where the positive and negative bands meet, forming a double cone, and (ii) four corresponding vHSs that occur at saddle points \emph{within} the BZ ($\kb=\half\kD$) where the density of states diverges logarithmically,
$g(\omega) \sim \log\abs{\omega-\omega_0}$
(Fig.~\ref{fig:1-pairing}F).~\cite{VanHove1953} Both features are hallmarks of quasiparticle spectra. 
 
 Remarkably, the periodicity of the spectrum allows one to define a smaller effective BZ (dashed black square in Fig.~\ref{fig:1-pairing}E top). This zone would be a primitive cell in a crystal with a doubled lattice constant, $2a$, another indication for pairing and quasiparticles. In this effective BZ, one can see the pairing mode as an optical phonon, with neighboring particles moving in opposite directions -- albeit, owing to the parity symmetry, the pairing modes have zero frequency unlike standard optical phonons. Taking into account all hydrodynamic interactions masks the pairing symmetry of the nearest-neighbor spectrum but preserves the topology of its critical points (Fig.~\ref{fig:1-pairing}E bottom), as verified in a simulation  (Fig.~\ref{fig:S2-spectrum}). 

The excitations at the Dirac points are pairing modes that generate lines of pairs (Fig.~\ref{fig:1-pairing}E). The Dirac cone describes long-wavelength acoustic modes of the pair lines. Due to the parity symmetry, the forces on particles in each pair are equal, and the pairing modes, $\kb=\kD$, are therefore marginally stable also when their amplitude is finite. In the nearest-neighbor spectrum, the Dirac and the acoustic cones are identical in shape, indicating the equal sound velocity of pairing modes and standard phonons (\ie the points $\Gamma$, X and M are identical). With all interactions taken into account, the cones are flattened towards the center, slowing down the propagation of modes in the $\Gamma$-X direction. 

For their purely real frequencies, the phonons excited in the linear dynamics (Eqs.~(\ref{eq:schrodinger},\ref{eq:Omegas})) are marginally stable. Hence, any instability or damping can only stem from non-linear coupling of the phonons. To examine this possibility, we followed the progression of the power spectral density (PSD) in a simulation starting with a white noise (Fig.~\ref{fig:1-pairing}G). Evolution of the PSD indicates strong selection of low-frequency excitations, presumably due to multi-phonon scattering events,\cite{Ziman1960} with sharp peaks at the Dirac pairing modes. \\

\begin{figure*}[!htb]
\centering
\includegraphics[width=0.95\textwidth]
{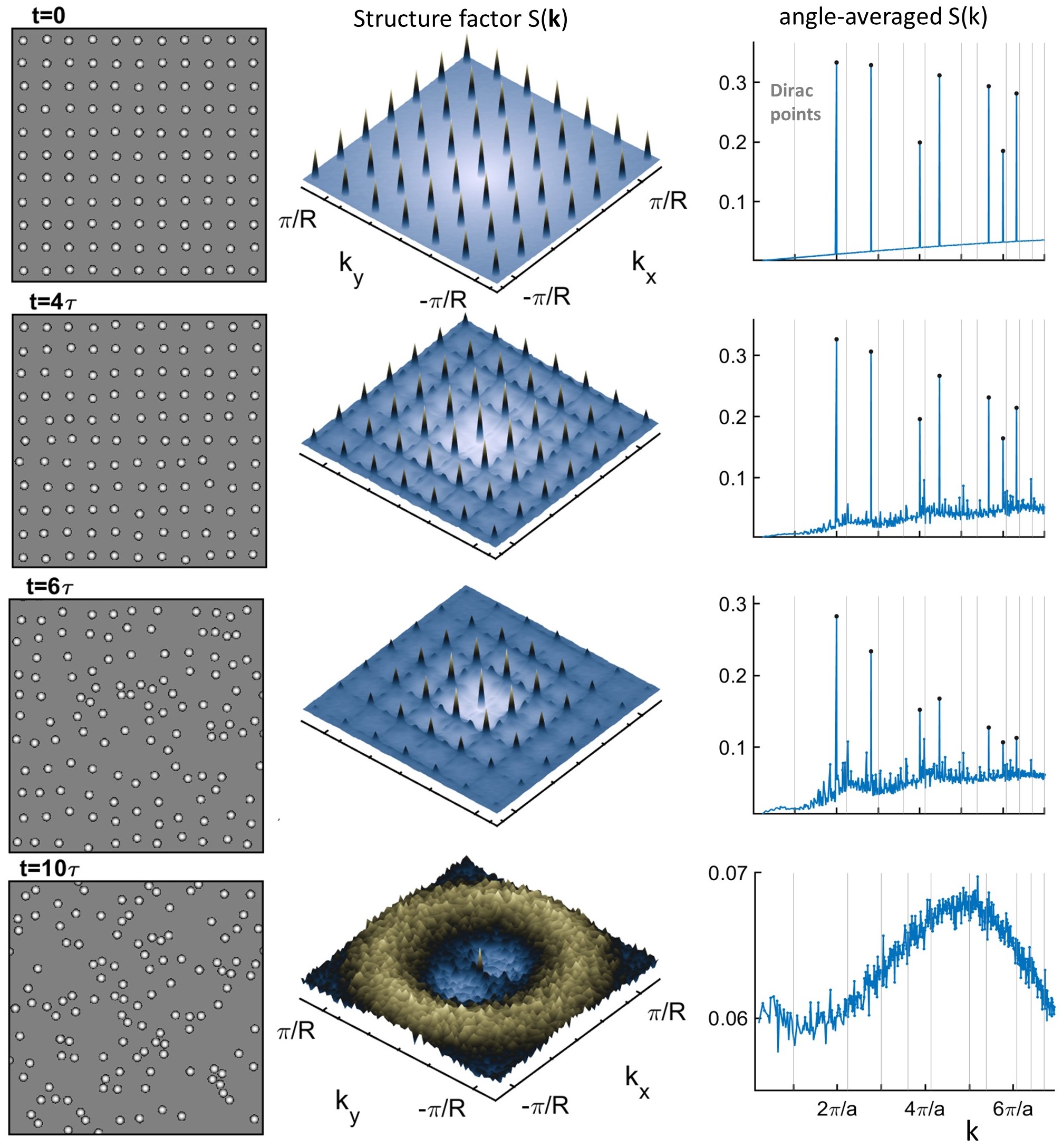}
\caption{
\sb{Pair-induced melting in the square lattice.}
Progression of the configuration in a simulated  $51 \times 51$ square lattice with lattice constant $a=5R$ at $t = 0, 2, 6, {\rm and~} 10 \tau$.
Left: Real space configuration.
Middle: The 2D structure factor $\Sk$. The Dirac peaks emerge in the X midpoints between Bragg peaks. 
Right: angle-averaged structure factor $S(k)$ showing positions of Dirac points (vertical grey lines) and Bragg peaks (black dots). The progression is shown in \href{\movieTwoSqMelt}{Movie 2}.
}
\label{fig:2-square}
\end{figure*}
\begin{figure}[!htb]
\centering
\includegraphics[width=1.00\columnwidth]{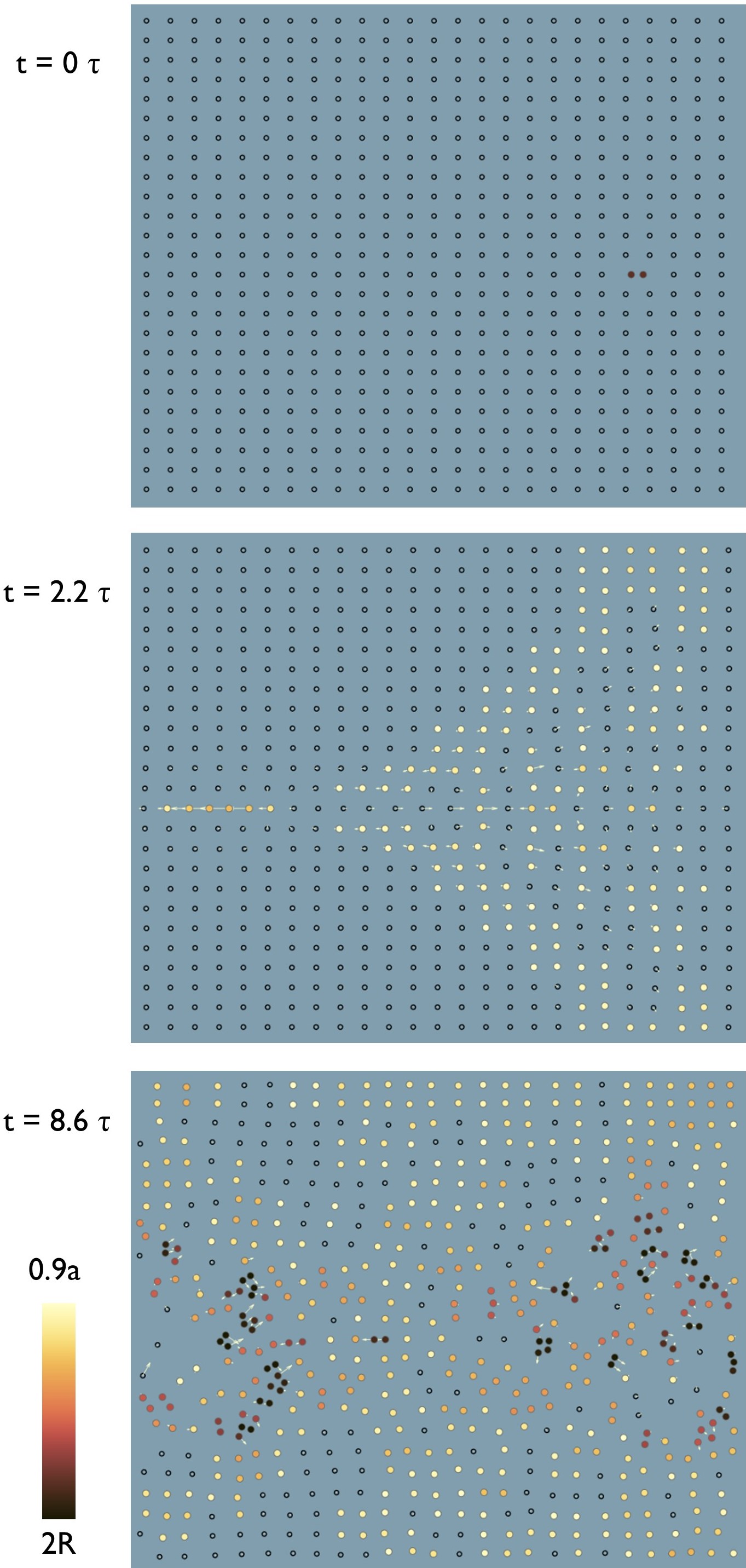}
\caption{
\sb{Quasiparticle avalanche.}
Top: A simulation starting from a perfect square lattice doped with an isolated pair quasiparticle ,positioned at the right-center (\href{\movieThreeAvalanch}{Movie 3}). Middle: The pair is propagating to the left while exciting an avalanche of pairs in a trailing Mach cone. Bottom: Collisions among the excited pairs induce melting (bottom). White arrows denote velocity and the distance to the nearest neighbour (\ie pair length) is color-coded between $2R-0.9a$. 
 follows the progression.}
\label{fig:3-avalanche}
\end{figure}

\begin{figure*}[!htb]
\centering
\includegraphics[width=1.00\textwidth]
{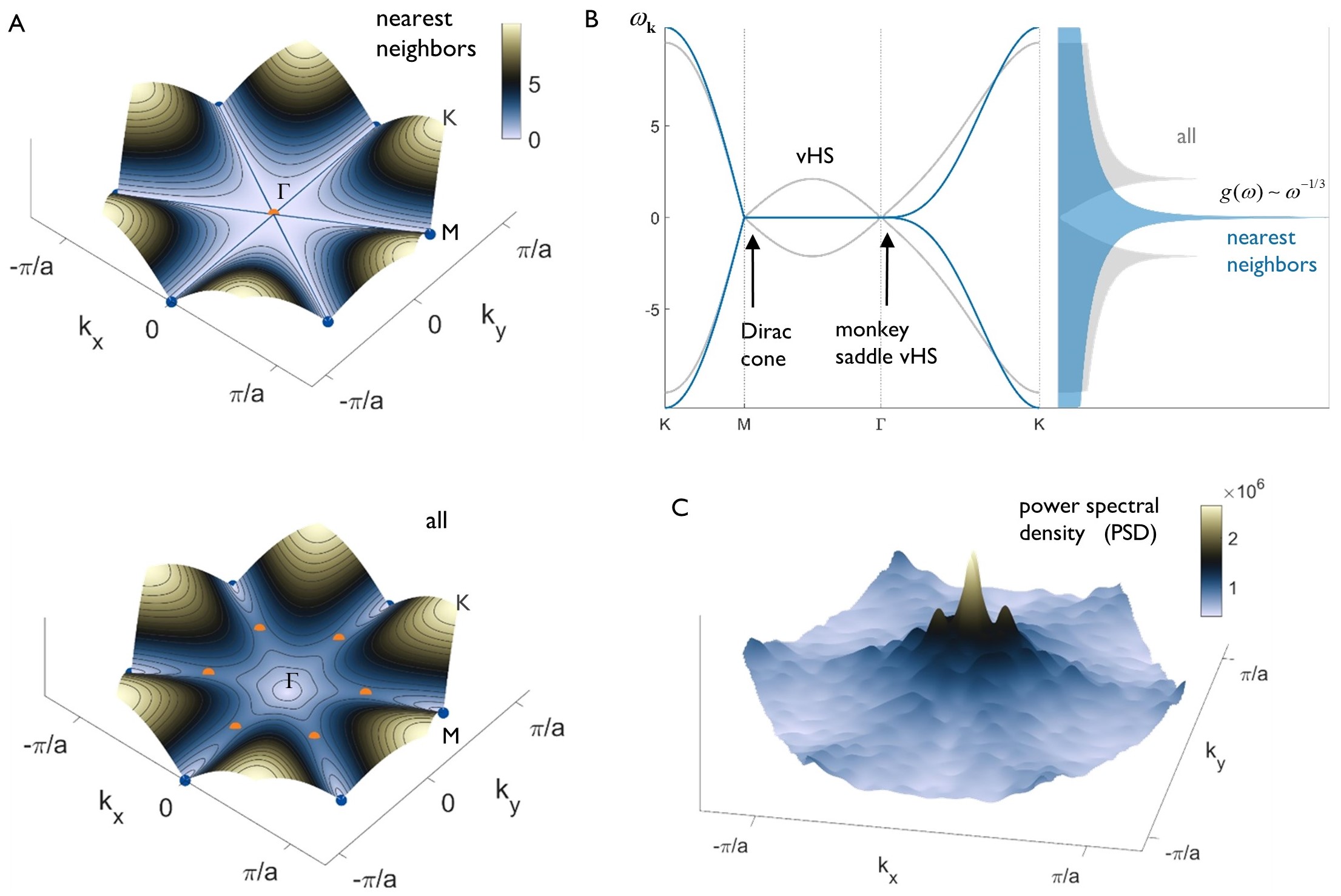}
\caption{
\sb{Flat bands and monkey saddles in hexagonal crystals.}
\sb{A.} Top: The frequency bands $\omk$ of a hexagonal lattice with nearest neighbor interactions, in the first Brillouin BZ. The six Dirac points (blue) extend into 1D zero-frequency lines (solid blue) connected to a ``monkey saddle" vHS at the center (orange point), where the band is nearly flat, $\omk \sim \pm k^3 \abs{\cos 3 \phi}$.
Bottom: the spectrum including all interactions with six standard vHSs (orange) and six Dirac cones (blue). 
\sb{B.}~The band structure and corresponding density of states, showing power-law divergence in the nearly-flat band, $g(\omega) \sim \omega^{-1/3}$.
\sb{C.}~PSD in a simulation starting from a white-noise perturbation shows selection of low-frequency modes in the flat band around $\kb = 0$. 
}
\label{fig:4-hex_spec}
\end{figure*}

\noindent\sb{Pair-induced melting}\\
The observation of pairing phenomena in both disordered and ordered phases puts forward a possible role of these excitations in the emergence of disorder. To examine this hypothesis, we performed numerical simulations, starting from an ordered square crystal (with little white noise) and following the progression of its structure and dynamics (Fig.~\ref{fig:2-square} and 
\href{\movieTwoSqMelt}{Movie 2}). At each time step, the structure factor $\Sk$, the squared Fourier transform of the configuration, and its angular-averaged form $S(k) = \lang \Sk \rang_{\abs{\kb} = k}$, are evaluated (the corresponding pair correlation function $g(r)$ is shown in  Fig.~\ref{fig:S3-gr} left). 
In crystals, the natural timescale is $\tau = a^3/(u R^2)$, the typical time it takes a perturbation to propagate a distance $a$.

At first, only Bragg peaks are noticeable in the structure factors $\Sk$ and $S(k)$.  After a typical time of a few $\tau$, peaks emerge at the Dirac points (the midpoints X between the Bragg peaks), which correspond to the acoustic pairing spectrum (Fig.~\ref{fig:2-square}). The emergence of these Dirac peaks concurs with the appearance of a ring-shaped modulation in $\Sk$ and $S(k)$. As the melting progresses, this annular pattern reveals itself as the structure factor of the disordered system, fittingly peaked at  $k = 2\pi/\ell$, corresponding to the particles' hard core. 

To gain further insight into the path to melting, we simulated the dynamics of a perfect crystal doped with a single defect -- an isolated quasiparticle (Fig.~\ref{fig:3-avalanche} and \href{\movieThreeAvalanch}{Movie 3}). 
This Dirac quasiparticle is coasting horizontally while exciting an avalanche of new quasiparticles, mostly arranged in pairing waves (\ie Dirac phonons with $\kb=\kD$, Fig.~\ref{fig:1-pairing}E). This ``Mach cone" of pairing is trailing behind the original quasiparticle that traverses the crystal supersonically. This is because the quasiparticle is a finite-amplitude disturbance, which moves faster than sound (whose velocity is the Dirac velocity, the slope of the Dirac cone). The ${\sim}$\SI{45}{\degree} angle of the cone indicates a Mach number of ${\sim}1.4$. Stripes of compression (pairing) and rarefaction waves are noticeable inside the cone. After about $\SI{6}{\tau}$, a pair-rich band stretches along the crystal. Then, collisions among pairs and phonons eventually lead to the breakup of any remaining crystalline order and the emergence of a fully-developed random phase. This melting process is driven by anharmonic terms in the equations of motion, beyond the linear Schr\"{o}dinger equation (Eq.~(\ref{eq:schrodinger})). The emergence of Dirac peaks in $\Sk$ and quasiparticle avalanches reveals pairing as the mechanism inducing the non-equilibrium melting transition.\\  
%

\begin{figure*}[!htb]
\centering
\includegraphics[width=0.95\textwidth]{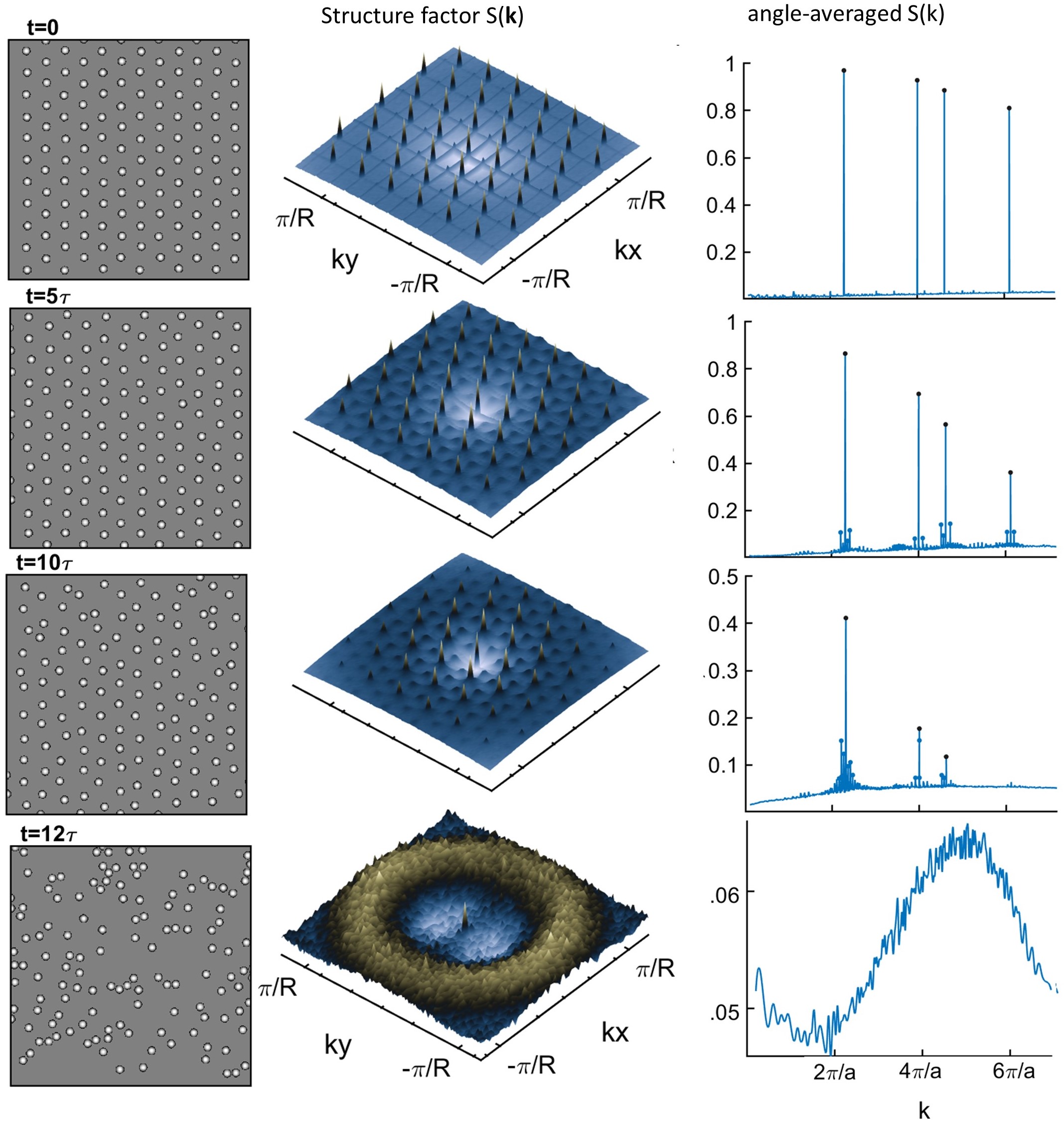}
\caption{
\sb{Melting in the hexagonal lattice.}
Simulation of a hexagonal lattice with lattice constant $a = 5R$ at $t = 0, 5, 10, 12 \tau$.
Left: Real space configuration.
Middle: The 2D structure factor $\Sk$. Flat bands appear as widened peaks around the Bragg points.
Right: angle-averaged structure factor $S(k)$ with Bragg peaks denoted by black dots. Notable are the emergent flat band modes, especially around the first peak.
\href{\movieFourHexMelt}{Movie 4} follows the progression.}
\label{fig:5-hex_melt}
\end{figure*}

\noindent\sb{Flat bands and monkey saddles in hexagonal crystals}\\
Hexagonal crystals are unique as the only class of 2D Bravais lattices whose symmetry matches the intrinsic threefold symmetry of the hydrodynamic interaction (the ``spring constant", Eq.~(\ref{eq:Omegas})), bringing about a qualitatively different pathway to disorder. A first hint comes from observing the nearest-neighbor spectrum of the hexagonal crystal,
$\omk \sim \pm \abs{\sin{\half k_x a}} \abs*{\cos{\half k_x a} - \cos{\halfsqthree k_y a}}$, which exhibits a remarkable pattern of critical points (Fig.~\ref{fig:4-hex_spec}A): (i) a single vHS is positioned exactly at the $\kb = 0$ center of the BZ (point $\Gamma$), and (ii) the Dirac points (M) extend into a web of zero-frequency lines, connecting the vHSs. These Dirac cones are flattened into ``wedges" (or ``canyons") stretched along the $\Gamma$-M direction. 

Importantly, this vHS of the hexagonal crystal is a``monkey saddle",\cite{Efremov2019}
a multicritical Lifshitz point where three canonical vHSs fuse into an elliptical umbilic catastrophe.~\cite{Shtyk2017} The long-wavelength expansion of dispersion around this vHS is $\omk \sim \pm k^3 \abs{\cos 3 \phi}$ (in polar coordinates $\kb = (k,\phi)$), representing two interlacing monkey saddles (Fig.~\ref{fig:4-hex_spec}B). The physical significance of the monkey saddle is the formation of a nearly-flat band with vanishing group velocity and curvature, $\partial_\kb\omega = \partial^2_{\kb}\omega = 0$.
The outcome is a power-law divergence of the density of states, 
\begin{equation}
    g(\omega) \sim \omega^{-\frac{1}{3}}~,
\end{equation}
much stronger than the logarithmic divergence at canonical vHSs (Methods). Such extreme slowing down of the excitations occurring in flat bands, multicritical and extended vHSs is known to induce strong correlations, and was proposed as a mechanism underlying high-$T_c$ superconductivity.~\cite{Gofron1994,McChesney2010,Kopnin2011,Shtyk2017,Isobe2018,Marchenko2018}

With all long-range interactions included, the symmetry of the double monkey saddle is broken, as it splits into six canonical vHSs, and the Dirac cones regain their standard shape. Nevertheless, the band remains relatively shallow in the $\Gamma$-M direction (Fig.~\ref{fig:4-hex_spec}AB), as verified in the simulation (Fig.~\ref{fig:S2-spectrum}). The evolution of the power spectral density (PSD) in a simulation starting with a white-noise exhibits strong amplification of slow excitations in the flat band around the saddle-monkey vHS (Fig.~\ref{fig:4-hex_spec}C).

Following the the progression of a hexagonal hydrodynamic lattice, we see a melting transition governed by the flat band. The dominant modes that appear in the structure factor $\Sk$ at the time of the melting transition are long-wavelength excitations sitting in the monkey saddle around each Bragg peak (Fig.~\ref{fig:5-hex_melt} and \href{\movieFourHexMelt}{Movie 4}). This flat band spectrum is amplified as the system approaches the melting transition, as manifested in the widening peaks (most notably at $t \simeq 10 \tau$).  As in the square lattice, a radial modulation emerges and eventually becomes the structure factor of the disordered phase.

\begin{figure}[!htb]
\centering
\includegraphics[width=1.00\columnwidth]
{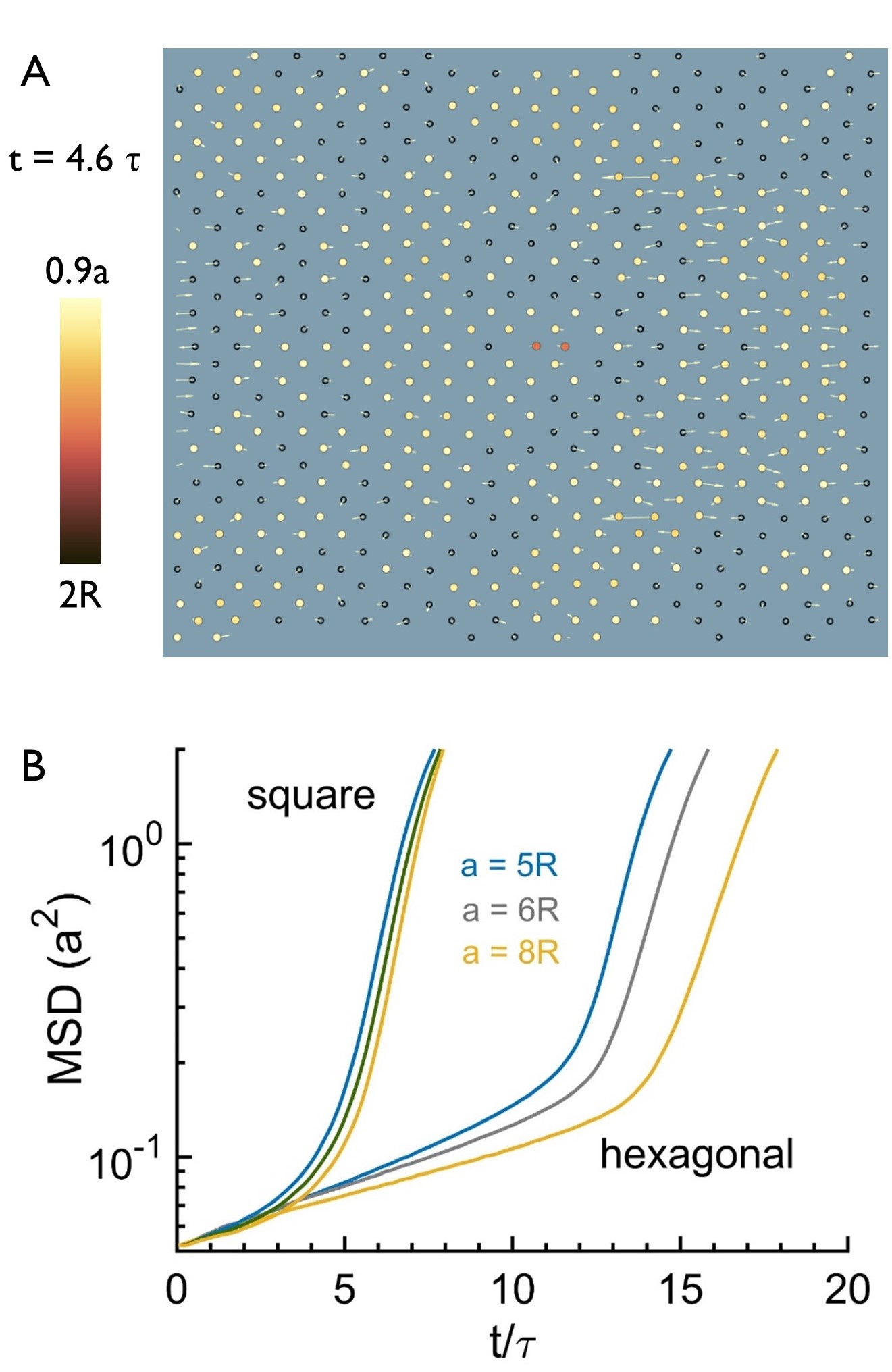}
\caption{
\sb{A}: A simulation of a hexagonal lattice doped with an isolated pair quasiparticle (orange particles). The pair excites ultra-slow flat-band phonons (white and yellow regions) while remaining practically static for an extended period, concluded by an abrupt melting transition (see \href{\movieFiveFlatband}{Movie 5}). Pair length is color-coded between $2R-0.9a$. 
\sb{B}: Progression of the mean squared deviation (MSD) in square (simulation in Fig.~\ref{fig:2-square}) and hexagonal (Fig.~\ref{fig:5-hex_melt}) crystals for $a/R = 5,6,8$, where time is measured in units of $\tau = a^3/(u\ell^2)$.  }
\label{fig:6}
\end{figure}

To further examine the role of the flat band, we follow the evolution of a hexagonal crystal doped with a single pair (Fig.~\ref{fig:6}B and \href{\movieFiveFlatband}{Movie 5}). Unlike the quasiparticle avalanche in the square crystal---here, the pair remains wobbling around its original position, surrounded by a sea of excited flat-band phonons, for an extended period of ${\sim}\SI{20}{\tau}$. The quasiparticle stays put due to the ultra-slow group velocity in the flat band, $\partial_\kb{\omega} \simeq 0$ (Fig.~\ref{fig:4-hex_spec}A,B). Then, many pairs rapidly emerge, presumably via multi-phonon collisions, inducing a swift melting of the crystal. 

The more collective nature of the transition also shows in a sharp change in the slope of exponential growth of the disorder parameter, the mean squared deviation (MSD) from the crystal positions (Fig.~\ref{fig:6}B). In comparison, the MSD of the square lattice grows continuously and super-exponentially. The square crystal MSD curves overlap when scaled by $\tau \sim a^3$. In contrast, the hexagonal MSD curves overlap when normalized by a timescale ${\sim}a^{7/2}$ (Fig.~\ref{fig:S4-MSDhex}), another manifestation of the dissimilar nature of these two melting transitions.  \\

\noindent\sb{Discussion and outlook}\\
\noindent The present findings demonstrate that quantum matter concepts---quasiparticles, van Hove singularities, and flat bands---provide insight into the many-body dynamics of a classical dissipative system.
It is instructive to consider the similarities and dissimilarities to graphene. The hydrodynamic interactions in the flowing crystal yield Dirac cones as in graphene. However, the cones are the outcome of the \emph{intrinsic parity symmetry} of the hydrodynamic force, $\fb(-\rb) = \fb(\rb)$, whereas in graphene, the cones result from the crystal symmetry of the honeycomb lattice, which consists of two interpenetrating hexagonal lattices. Thus, the analog of the graphene pseudo-spin is the polarization vector of the excitations $\psik$. The different underlying symmetries give rise to distinct spectra. In graphene, the vHSs occur on the boundary of the BZ, at the M points, and the Dirac cones are at the K corners.~\cite{CastroNeto2009}

Exact solutions for the steady-state motion of a pair of finite-size spheres were found in the 20\textsuperscript{th} century,~\cite{Stimson1926,Goldman1966} (the 21\textsuperscript{st} for a quasi-2D geometry.~\cite{Beatus2012,Sarig2016}) But it is tempting to assume that the stability of the pair, a direct outcome from the symmetry of the Stokes equation, was already known to Stokes himself. So in hindsight, the hydrodynamic pairs are very old classical quasiparticles.
In classical condensed matter, the collision time is typically too short to allow long-lived particle-like excitations, such as the quasiparticles observed in low-temperature quantum matter.~\cite{Woelfle2018} Nevertheless, the present system generates a macroscopic fraction of such particle excitations in the form of hydrodynamic pairs. The pairs are stable enough to be seen as compound particles with a well-defined velocity because, effectively, the system is a dilute ensemble of dipole-like particles, even though their interactions are mediated by the surrounding dense fluid.

The nearly-flat band and the multicritical vHS exhibited in hydrodynamic crystals with threefold symmetry are of special interest in many-body physics. As demonstrated in the present system, the excitations in the flat band become extremely slow, leading to strong correlations and collective modes, which give rise to a sharper melting transition. Flat bands were recently found in bilayer graphene twisted at a specific magic angle~\cite{Bistritzer2011} or buckled.~\cite{Mao2020} We observed similar divergences in the density of excitations of a driven hydrodynamic system, raising the possibility that other emergent many-body phenomena of 2D electronic systems may be revealed in classical dissipative settings. 
As for future directions, the present findings propose that quantum matter notions can be widely useful for examining emergent many-body phenomena, particularly non-equilibrium phase transitions, in a variety of classical dissipative systems, ranging from soft matter, driven~\cite{Beatus2017,Tlusty2021} and active~\cite{Marchetti2013,Shankar2020} alike, and meta-materials~\cite{Kadic2019,Scheibner2020} to complex plasma,~\cite{Ivlev2015} reaction-diffusion,~\cite{You2020} chemotaxis,~\cite{Meredith2020} catalysts~\cite{Jee2018a,Wang2020}, and ecology~\cite{Sprott2009}.

\bibliography{NonHermitian}
\onecolumngrid

\clearpage
\twocolumngrid

\relscale{0.9}
{\larger[2] \sb{Methods}}\\

\sb{\larger[1] Experiment} \\
\noindent \sb{Setup.}~
We investigated the collective behavior of hydrodynamically interacting particles in a quasi-2D flow.
The particles were polystyrene microspheres of diameter \SIrange{7}{20}{\um} (Sigma-Aldrich). These microspheres were driven in water in microfluidic channels of height \SIrange{10}{30}{\um}. The typical roughness of the microscope cover-slips used as the floor and ceiling of the channel is $\le\SI{30}{\nm}$, not more \SIrange{0.04}{0.1}{\percent} of the channel height. The roughness induces quenched disorder, which locally perturbs the pressure and velocity fields by a similar magnitude and may generate pairs in a hydrodynamic crystal.~\cite{Tlusty2006} A stronger noise source is due to the polydispersity of the polystyrene beads, whose standard deviation is \SIrange{0.2}{0.3}{\um}.

The flow rate in the channel was controlled via a pressure pump (Fluigent LineUp). The motion of particles was recorded digitally at a rate of \SIrange{100}{2000}{\frames/\s} with a high-speed camera (Phantom V-1120) and tracked using MatLab software. To follow the motion of the ensemble, the camera is fixed on a translation stage, moving at the mean speed of particles. The areal density of particles in the field of view was \SIrange{0.3}{7}{\percent}.  One end of the channel was kept open to ensure that the particle flow remained steady and avoid pressure buildup in the channel.

The Reynolds and P{\'e}clet numbers for the flow were in the range $\Rey \simeq \numrange{e-4}{e-3}$ and $\Pec \simeq \numrange{e3}{e4}$, such that one can consider a low-Reynolds Stokes flow while disregarding any thermal and inertial effects. 
The typical velocities of the fluid and of the particles were in the range \SIrange{e2}{e3}{\um/\s}. The velocities of particles were calculated from the tracked trajectories. The particles are slower than the surrounding fluid (by about \SI{50}{\percent}) due to friction with the solid boundaries of the channel. This relative motion of the particles induces drag forces on the particles and perturbations in the flow field. \\

\noindent\sb{Tracking dynamics and pair statistics.}~
Pairs are defined geometrically as couples of particles that are much closer to each other than to other particles. Thus, the forces they exert on each other are much stronger than interactions with other particles, and they can be seen as compound, weakly-interacting particles.  Specifically, a particle belongs to a pair if the distance to the nearest neighbor (\ie the other particle in the pair) $r_1$ is shorter by a factor of \numrange{3}{4} than the distance to the next-nearest particle, $r_2$. In the analysis of the experiment and simulations, we take $3.5 \, r_1 \le r_2$. Due to the $1/r^2$ dependence, this ensures that the intra-pair forces are at least tenfold stronger than the interactions with other particles. 
The lifetime of a pair is computed from the auto-correlation function of the distance $r_1(t)$. The lifetime (measured in units of $R/u$) is defined when the correlation decreases by a factor of \SI{25}{\percent}. \\

\sb{\larger Model}\\
\noindent\sb{Hydrodynamic forces.}~
 Consider particles of size $\ell = 2 R$ are moving in the $x$-$y$ plane of a thin 2D fluid layer between rigid floor and ceiling (a Hele-Shaw cell), at a velocity $u$ relative to the fluid. 
 In this quasi-2D setting, the narrow dimension is $z$ (perpendicular to the page in Fig.~\ref{fig:1-pairing}). The viscous drag on each particle is $\gamma \ub$, where $\gamma$ is the friction coefficient. All particles are driven by a force $\Fb$ along the $x$-axis, which compensates for the viscous friction. In the experiment, $\Fb$ is the friction of the particles with the floor and ceiling.
The particles' motion with respect to the fluid induces dipolar perturbations with a velocity field decaying as ${\sim} u \, (R/r)^2$, and these dipolar flow fields give rise to collective hydrodynamic interactions \cite{Beatus2006, Baron2008, Beatus2012, Shani2014,Beatus2017}. The hydrodynamic force $\fbij$ exerted by the \jth particle on the \ith particle is (in $x,y$ components),~\cite{Tlusty2021}
\begin{equation}
    \fbij  = \fb(\rbij) =
    \alpha \cdot \gamma u \l( \frac{R}{\rij} \r)^2
    \begin{bmatrix}
    \cos{2 \tetij} \\ \sin{2 \tetij}
    \end{bmatrix}~,
    \label{eq:fhyd-M}
\end{equation}
where  the positions of the dipoles are $\rb_i$, and $\rbij =\rb_i -\rb_j$ are the distance vectors. In polar coordinates, $\rbij = (\rij,\tetij)$, where $\rij = |\rbij|$ and $\tetij$ the angle. The geometric factor $\alpha\sim o(1)$ depends on the shape of the particles (\eg disks, spheres). 
Equation (\ref{eq:fhyd-M}) implies that $\fbji  =  \fbij$, since the angles obey $\tetji = \pi + \tetij$. This implies that an isolated pair moves at a uniform velocity $\ubp = \fb(\rbij)/\gamma = \alpha u \cdot (R/r)^2 (\cos{2\theta}, \sin{2\theta})$ (Fig.~\ref{fig:1-pairing}B), where in the experiment $\alpha \simeq \numrange{0.3}{0.5}$.

We see that the hydrodynamic forces violate Newton's third law of momentum conservation. While the microscopic molecular forces in the fluid obey Newton's law, the hydrodynamic interactions are \textit{effective macroscopic} forces that do not conserve momentum. This is because the viscous flow is an inherently open system, an effective representation of energy and momentum transfer from hydrodynamic degrees of freedom to microscopic ones. For a detailed treatise on the physics of 2D microfluidic ensembles, see Beatus \textit{et al.} \cite{Beatus2012,Beatus2017}\\

\noindent\sb{Equations of motion.}~
The hydrodynamically-interacting particle ensembles exhibit complicated chaotic dynamics in the fully-disordered phase and non-linear mode-coupling in the ordered, crystalline phase. To compute their trajectories, one could in principle solve the underlying Stokes equations consistently with the moving boundaries of the particles, albeit this is in general a rather cumbersome procedure. One possible approach is expanding the hydrodynamic interactions as a multipole series over ``hydrodynamic image charges" induced by the particles' solid boundaries.~\cite{Pozrikidis1992} The procedure is very similar to electrostatics since the same Laplace equation solves the Hele-Shaw flow potential. For example, if we consider two finite particles of size $\ell$, then the first image in the interaction will be a dipole ${\sim}(\ell/r)^2$, followed by an infinite series of multiple reflections, $(\ell/r)^4, (\ell/r)^6,\ldots$ Thus, in an ensemble of particles, one in principle needs to sum over all possible multiple scattering paths among all the particles. Fortunately, since the system is always dilute ($\le\SI{7}{\percent}$ areal density), we can neglect all the higher terms and take only the first reflections. This is an excellent approximation, validated in numerous studies.~\cite{Cui2004,Beatus2006,Beatus2007,Beatus2008,Beatus2009,Champagne2011,Liu2012,Desreumaux2013,Shani2014} For a detailed discussion of the hydrodynamic images sum and their convergence, see Beatus \textit{et al.}~\cite{Beatus2012,Beatus2017}

Thus, we can use the following method, relying on two well-established approximations: (i) unless the particles almost touch each other, their induced velocity perturbations are approximated by the isolated dipole field (Eq.~(\ref{eq:fhyd-M})); (ii) the total hydrodynamic force acting on a particle can be simply computed as the sum over the \textit{pairwise} interactions with all other particles.~\cite{Beatus2006,Beatus2012,Beatus2017} In the low-Reynolds regime, inertia is negligible, so the drag force is balanced by the driving force and the hydrodynamic interactions. The resulting system of $N$ coupled equations of motion is
\begin{equation}
    \gamma \dot{\rb}_i =\Fb + \sum_{j\neq i}{\fb}\l(\rb_i-\rb_j\r) = \Fb + \FF_i\l(\{ \rb_j \}\r)~,
    \label{eq:M-motion}
\end{equation} 
where $\FF_i$ is the overall hydrodynamic force acting on particle $i$. To sum, in these coarse-grained equations, the hydrodynamic degrees-of-freedom are ``integrated out", \ie we do not calculate explicitly the velocity and pressure fields. Technically speaking, we replace the integration of a linear PDE with $N$ moving boundaries, by a nonlinear system of ODEs for the trajectories of these boundaries (Eq.~(\ref{eq:M-motion})), equivalent to the simulated system in Eq.~(\ref{eq:M-simulation}).\\

\noindent\sb{Dynamic equation of moving crystals}.~
At steady-state, the lattice interactions in Eq.~(\ref{eq:M-motion}) vanish by symmetry, $\FF_i = \sum_{j \neq i}{\fbij} = 0$, and the lattice moves uniformly at a velocity $\ub = \Fb/\gamma$, relative to the surrounding fluid. Expansion of the equations of motion in small deviations of the lattice positions around the steady-state positions, $\delrb_j = \rb_j - \bar{\rb}_j$, yields a linear dynamic equation, 
\begin{equation}
\label{eq:M-dynamic}
    \dot{\delrb} = \HH \, \delrb~,
\end{equation}
where $\delrb$ is $2N$-vector of the $N$ particle deviations. 
The tensor $\HH^{ij}_{\alpha\beta} = \pdv*{\FF_{i\alpha}}{\rb_{j\beta}}$ is a generalized \emph{spring constant} that multiplies the deviation $\delrb$ to give the hydrodynamic force (where $\alpha,\beta$ are the coordinates $x,y$). $\HH$ is a $2N \times 2N$-matrix composed of $2 \times 2$ blocks $\Hij$ that account for hydrodynamic interactions between the \ith and \jth particles, 
\begin{equation*}
    \Hij= H(\rb_i-\rb_j) \equiv \pdv{\FF_i}{\rb_j}= 
    \frac{2}{\brij^{3}}
    \begin{bmatrix} 
    \cos{3 \tetij} & \sin{3 \tetij}\\ 
    \sin{3 \tetij} & -\cos{3 \tetij}
    \end{bmatrix}~.
\end{equation*}
The diagonal terms ensure zero sums, $\HH^{ii} = -\sum_{j \neq i}{\Hij}$. The three-fold symmetry of $\HH$ stems from its definition as the derivative of the force, $\HH=\fdv*{\fb}{\rb}$ whose symmetry is two-fold.  Note that $\HH$ is translation invariant ($\HH^{ij}$ is a function of only $\rb_j - \rb_i$) and anti-symmetric, $\HH^{ji}= -\HH^{ij}$. In a crystal, the diagonal elements vanish due to reflection symmetry, $\HH^{ii}=0$. Thus, $\HH$ of a crystal is skew-Hermitian with $N$ purely imaginary eigenvalues, representing $N$ phononic modes.
                                       
Hereafter, we measure the physical quantities by the relevant scales of the crystal: Distances are measured in units of $a$, the typical distance between the particles (and wavevectors in $1/a$). In a lattice, $a$ is the lattice constant.  Times are measured in units of $\tau$, the timescale of the hydrodynamic interaction, $\tau \equiv  a^3/\l(u \ell^2 \r)$, the time it takes a perturbation to traverse a distance $a$ (frequencies are measured in $1/\tau$). \\

\noindent\sb{Momentum space}.~
The equations of motion are then expanded in plane waves with 2D wave-vectors $\kb =(k_x,k_y)$, such that the deviation $\delrb$ of each particle from its mechanical equilibrium position $\bar{\rb}_j$ is
\begin{equation*}
    \delrb_j(t) = \Psik(t) \, e^{i \kb \cdot \brbj} = 
    \psik e^{i \l( \kb \cdot \brbj- \omk t \r)}~.
\end{equation*}
Here, $\Psik(t) = \psik e^{-i \omk t}$ is a 2D polarization vector in $\kb$-space. $\Psik(t)$ evolves according to a Schr\"odinger-like equation,
\begin{equation*}
    i \pdv{t}\Psik(t) = \Hk \Psik(t)~,
\end{equation*}
with an eigenvector $\psik$ and eigenfrequency $\omk$ obeying Eq.~(\ref{eq:schrodinger}),
\begin{equation*}
        \Hk \psik = \omk \psik~.
\end{equation*}
The ``Hamiltonian" $\Hk$ is a $2\times 2$-matrix, which is a Fourier-transform of $\HH^{ij}=H(\brb_j-\brb_i)$,
\begin{equation*}
    \Hk = i \sum_j{H(\brb_j-\brb_i) \, e^{i\kb\cdot (\brb_j-\brb_i)}}~,
\end{equation*} 
where we multiplied by the imaginary unit for convenience, such that $\Hk$ becomes Hermitian. There are $N$ operators $\Hk$ (one for each $k$) with $N$ real phonon eigenfrequencies $\omk$. Owing to the translation symmetry of both the crystal and $\HH^{ij}$, $\Hk$ is also translation-invariant.
$\Hk$ can be expressed in terms of Pauli's matrices as
\begin{equation}
    \Hk = \Omx \sigz + \Omy \sigx~.  
    \label{eq:Hk-M}
\end{equation}
The contributions of the long-range hydrodynamic interaction to $\Hk$ are Fourier sums,
\begin{equation}
     \Omegab_\kb = 
     \begin{bmatrix}
    \Omx \\ \Omy
    \end{bmatrix} =  
    \sum_{ j \neq 0 } 
    \frac{2}{\brj^{3}}
    \begin{bmatrix}
    \cos{3 \tetj} \\  \sin{3 \tetj}
    \end{bmatrix}
    \,\sin{\l(\kb \cdot \brbj\r)} ~,  
    \label{eq:Omegas-M}
\end{equation}
where $\brbj = \br_j(\cos{\tetj},\sin{\tetj})$ are the distances of the steady-state lattice positions from an arbitrary origin particle $0$. Due to the crystal's parity symmetry, $\Omx$ and $\Omy$ are always real. 
Since $\Omx$ and $\Omy$ are odd functions of $\kb$, $\Hk$ is also odd under parity, $\mathcal{H}_{-\kb} =  - \Hk$.
One can represent the Hamiltonian in the basis of left and right circularly polarized unit vectors, $2^{-1/2}[1, \pm i]^\T$ as $\Hk  = \Omegab_\kb \cdot \sigmab = \Omx \sigx + \Omy \sigy$ (Eq.~(\ref{eq:schrodinger})). \\

\noindent\sb{Spectra}.~
The eigenfrequencies $\omk$ are found by solving the secular equation corresponding to Eq.~(\ref{eq:Hk-M}).  There are two eigenfrequency bands, $\omk = \pm \abs{\Omegab_\kb} =\pm ( \Omx ^2 + \Omy^2)^{1/2}$, and the corresponding polarization eignevectors are 
\begin{equation*}
      \psik^{+} = \frac{1}{\sqrt{2}}
      \begin{bmatrix}
      \cos{\frac{\alpha}{2}}\\ \sin{\frac{\alpha}{2}}
      \end{bmatrix}~,~
      \psik^{-} = \frac{1}{\sqrt{2}}
      \begin{bmatrix}
      -\sin{\frac{\alpha}{2}} \\ \quad \cos{\frac{\alpha}{2}}
      \end{bmatrix}~,
\end{equation*}
where the angle $\alpha = \arg{ \l(\Omx + i \Omy\r) } = \arctan{\l(\Omy/\Omx\r)}$.
In the circular basis, the eigenvectors are
\begin{equation*}
      \psik^{\pm} = \frac{1}{\sqrt{2}}
      \begin{bmatrix}
     \pm e^{ \pm i \frac{\alpha}{2}}\\ 
     \quad  e^{ \mp i \frac{\alpha}{2}}
      \end{bmatrix}. 
\end{equation*} \\

\noindent\sb{Nearest-neighbor approximation}.~ For the sake of simplicity, we consider the case where the particles interact only with their nearest neighbors. For a square crystal of lattice constant $a$, one finds from summing Eq.~(\ref{eq:Omegas-M}) over nearest-neighbors the bands
\begin{align}
    \omk  & =  \pm 4 \sqrt{ \sin^2 k_x a + \sin^2 k_y a} 
    \\ \nonumber
    & = \pm 2^{3/2} \sqrt{ 2 - \cos{2 k_x a} - \cos{2 k_y a} }~.
    \label{eq:square-M}
\end{align}
In the first BZ, the inner vHS are located at the saddle points, $\kb = (\pm\half\pi,0)$, and $(0,\pm\half\pi)$. The Dirac points are the midpoints of the Brillouin zone edges, $\kb = (\pm\pi,0)$, and $(0,\pm\pi)$.
Likewise, for the hexagonal crystal, the bands are
\begin{equation}
    \omk = \pm 8 \abs{\sin{\half k_x a}} 
    \abs{\cos{\half k_x a} - \cos{\halfsqthree k_y a}}~.
    \label{eq:hexagonal-M}
\end{equation}
From Eq.~(\ref{eq:hexagonal-M}), we see that the hexagonal bands exhibit six $\omega = 0$ lines linking the Dirac cones and the vHS into a hexagonal web of flat bands. \\

\noindent\sb{Dirac points.}~
Dirac points occur at wavevectors $\kD$ for which the hydrodynamic interaction vanishes, $\Omx = \Omy = 0$. Equation (\ref{eq:Omegas-M}) implies this happens when $\sin(\kb \cdot \brbj) = 0$, \ie for $\kb$'s that are halves of the reciprocal space base vectors, $\half \bb_1, \half \bb_2$, and their combinations, $\kD = \half \beta_1 \bb_1 + \half \beta_2 \bb_2$ (where $\beta_1,\beta_2 \in \{-1,0,1 \}$). In the hexagonal lattice, the six Dirac points are $(\beta_1,\beta_2) = (0,\pm 1), (\pm 1,0), (\pm 1,\mp 1)$. Note that these are the M midpoints of the BZ edges and not the K corners as in graphene.

The expansion of Eq.~(\ref{eq:Omegas-M}) around the Dirac point is linear in $\dkb = \kb - \kD$, the wave-vector of the quasiparticles, $\Omx \sim \grad_\kb{\Omx}  \cdot  \dkb$ and $\Omy \sim \grad_\kb{\Omy} \cdot  \dkb$, where the gradients are the sums 
\begin{align*}
    \grad_\kb{\Omx}  \equiv 
    \begin{bmatrix}
    c_{xx} \\    c_{xy}
    \end{bmatrix} = 
    &  2  \sum_{ j \neq 0 } 
     \l( -1 \r)^{\beta_1 \alpha_1^j + \beta_2 \alpha_2^j }
     \begin{bmatrix}
     \cos{\tetj} \\
     \sin{\tetj}
     \end{bmatrix}
     \frac{\cos{3 \tetj}}{\brj^2}~,
     \label{eq:gradDP-M}\\
     \grad_\kb{\Omy} \equiv
     \begin{bmatrix}
     c_{yx} \\    c_{yy}
     \end{bmatrix} = 
     & 2 \sum_{ j \neq 0 } 
     \l( -1 \r)^{\beta_1 \alpha_1^j + \beta_2 \alpha_2^j }
     \begin{bmatrix}
     \cos{\tetj} \\
     \sin{\tetj}
     \end{bmatrix}
     \frac{\sin{3 \tetj}}{\brj^2}~. \nonumber
\end{align*}
Here, the $\alpha_1^j$ and $\alpha_2^j$ are the indices of the lattice positions, $\brbj = \alpha_1^j \ab_1 + \alpha_2^j \ab_2$, with the basis vectors, $\ab_1$ and $\ab_2$. The gradients at the Dirac point are orthogonal, $\grad_\kb{\Omx} \cdot \grad_\kb{\Omy}  = 0$, and the resulting cone is therefore elliptic (Fig.~\ref{fig:1-pairing}E,F),
$\omk^2 \simeq \l( c_{xx}^2 + c_{yx}^2\r) dk_x^2 + 
    \l( c_{xy}^2 + c_{yy}^2\r) dk_y^2$. \\

\noindent\sb{Density of state and its van Hove singularities.}~
The density of states $g(\omega)$ is calculated by numerical summation of the integral, 
\begin{equation}
    g(\omega) = \l( \frac{a}{2 \pi} \r)^2
    \int{ \dd[2]{\kb} \, \delta \l(\omega - \omega(\kb)\r)}~.
    \label{eq:dos}
\end{equation}
The vHSs~\cite{VanHove1953} are located at saddle points, where $\grad_\kb{\omega} = 0$ and the Gaussian curvature of the band, \ie the determinant of the Hessian, is negative, $\det \l( \partial_{k_\alpha k_\beta}\omega \r) < 0$. In 2D systems, the density of states diverges logarithmically at the vHS, $g(\omega) \sim \log \abs{\omega-\omega_0}$.\\

\noindent\sb{Monkey saddles.}~At higher-order multicritical vHSs, such as found in the hexagonal crystal, the Gaussian curvature also vanishes. The bands around this vHS are two interlacing ``monkey saddles", $\omk = \pm \half k^3 \abs{\cos 3 \phi}$, To calculate $g(\omega)$, we integrate along an isofrequency line. Around the monkey saddle, the norm of the gradient is $\abs{\grad_\kb{\omega}}\simeq (3/2) k^2$. We integrate in one sextant, $-\pi/6 \le \phi \le \pi/6$, and multiply by six. Along the isofrequency line, $k_\omega(\phi) \simeq (2 \omega/\cos 3\phi)^{1/3}$, the density of states Eq.~(\ref{eq:dos}) becomes
\begin{align}
        g(\omega) = & \frac{3}{2\pi^2} 
    \int{ \frac{\dd{k_\omega}}{\abs{\grad_\kb{\omega}}}} 
     \simeq \frac{ \Gamma\l(\sixth\r)}
    {18^{\frac{1}{3}}\pi^{\frac{3}{2}}\Gamma\l(\twothirds\r)} \times \omega^{-\frac{1}{3}}~. 
    \label{eq:dosmonkey}
\end{align}\\

\sb{\larger[1] Simulation}\\
As explained in the Model section of Methods, we solve the coarse-grained equations of motion (Eq.~(\ref{eq:M-motion})) for an ensemble of $N$ hydrodynamically interacting particles in quasi-2D geometry with hard-core repulsion. The corresponding system of $2N$ coupled ordinary differential equations  (for the $x$ and $y$ coordinates of each of the $N$ particles) are:
\begin{equation}
\label{eq:M-simulation}
    \dot{x}_i =  \alpha u\ell^2 \sum_{j \neq i} 
    \frac{\xij^2 - \yij^2}{(\xij^2+\yij^2)^2}~,~
    \dot{y}_i  =  \alpha u\ell^2 \sum_{j \neq i} 
    \frac{2\xij \yij}{(\xij^2+\yij^2)^2}~.
\end{equation} 
Here, the positions of the dipoles are $\rb_i=(x_i,y_i)$, and their velocities are $\dot{\rb}_i=(\dot{x}_i,\dot{y}_i)$, in the frame of reference of the lattice, and $\rbij = (\xij,\yij)$ are the distances. Using the natural space and time scales, $\ell$ and $\tau_0 = \ell/(\alpha u)$, the equations become dimensionless and invariant to the choice of $u$. Hard boundary conditions are imposed at the surface of particles to prevent any overlap. 

Equations (\ref{eq:M-simulation}) are numerically integrated, in short enough time steps, $\Delta t \le \SI{0.1}{\tau_0}$, yielding the trajectories of the particles. Typically, periodic boundary conditions are applied to avoid the boundary effects, but we also examined a finite system. Simulations are then repeated for a range of initial conditions, crystal or random arrangements, with a wide range of aerial density of particles, $\rho=\SIrange{0.3}{10}{\percent}$. 
In crystal simulations, a small level of white noise is usually imposed in the initial conditions, $|\delrb| \le \num{e-3} a$. Even without this initial imperfections, rounding errors in the simulation provide perturbations that eventually destabilize the crystalline state.\\

\noindent\sb{Pairing, structure factor, and correlation function.}~
The structure factor $\Sk$ is computed by transforming each frame of particle configuration into a high resolution image, where each particle is represented by a small circle of diameter ${\sim}0.2 R$ to avoid the effects of the form factor. Then, the squared modulus of the Fourier transform of the image yields $\Sk$. At the beginning of the simulation, only the Bragg peaks corresponding to the perfect crystal are apparent. For example, the peaks of the square lattice are at $\kb = (\pi/ a)\{2m,2n\}$, for all integers $m$ and $n$. As the dynamics progress, the amplitude of the Bragg peaks decreases, and other patterns emerge, most importantly peaks at the Dirac cones or flat bands. In the square lattice, the Dirac points are at $\kb = (\pi/ a)\{m,n\}$, where $m$ is odd and $n$ is even, or vice versa. These correspond to the radial positions $k=\abs{\kb}$ in $S(k)$ (Fig.~\ref{fig:2-square}):
\begin{equation*}
   k =\frac{\pi}{a}\cdot
   \{1, \sqrt{5},3,\sqrt{13},\sqrt{17},5,\sqrt{29},\sqrt{37},\sqrt{41},\sqrt{45},7,\ldots \}.
\end{equation*}
After the crystal is completely melted, all these peaks vanish, and $\Sk$ is dominated by the disordered state, exhibiting an annular peak around $k \simeq 2\pi/\ell$ (the position of the peak depends on the particle density). 
In this regime, the angle-averaged structure factor $S(k)$ is a Fourier transform of the radial pair correlation function, $g(r)$. 

 To calculate the radial pair correlation function $g(r)$, we count the number of particles, $\dd{n(r)}$,  within an annular region $2\pi r \dd{r}$ around each particle in the ensemble, where periodic boundary conditions are employed to avoid finite size effect. Then, the pair correlation function is computed as a double average, over all particles in the system in multiple simulations, which is normalized by the number of particles in an uncorrelated system, $g(r) = \lang\dd{n(r)}\rang/(2\pi r \dd{r} \cdot \rho)$.\\
 
\noindent\sb{Dispersion relation and power spectral density.}~
The positions of all particles are recorded at each time step. Then, the deviations from crystal positions or the perturbations of the distances between nearest neighbors are calculated in the square and hexagonal geometries. The data is Fourier transformed in time and space, yielding two three-dimensional arrays (2D in $\kb$ and 1D in $\omega$) for perturbations in the $x$ and $y$ directions, $X_{\kb,\omega}$ and $Y_{\kb,\omega}$. The calculation is carried out until the perturbation is too large and the original order of the particles is lost. The dispersion $\omk$ and the power spectral density ($\PSD$) are obtained from combining peaks of $\abs{X_{\kb,\omega}}^2$ and $\abs{Y_{\kb,\omega}}^2$ at each point in the $\kb$-plane. \\

\noindent\sb{\larger[1]Data Availability}\\
\noindent Data supporting the figures within this paper are available from the corresponding authors upon reasonable request.\\

\noindent \sb{\larger[1]Code Availability}\\
The code used for the analysis of the experiment, analytic modeling, and simulations in this study is available from the corresponding authors upon reasonable request.\\

\noindent \sb{\larger[1]Acknowledgements}\\
This work was supported by the Institute for Basic Science, Project Code IBS-R020. We thank Issac Michael and Yoon-Kyoung Cho for their essential help in constructing the microfluidic channels. T.T. thanks Sam Safran for crucial comments on quasiparticle spectra, Roy Bar-Ziv and Tsevi Beatus for many fruitful discussions, and Elisha Moses for comments on the manuscript. We are grateful to the anonymous reviewer for deep remarks on the underlying physics.\\

\noindent \sb{\larger[1]Author Contributions}\\
\noindent I.S. performed the experiment, analyzed the measurements and ran simulations. H.K.P. and T.T. designed and supervised the research. T.T. initiated the study of quasiparticles and flat bands in soft matter, and developed the physical theory. H.K.P., I.S. and T.T. conducted the research, wrote and revised the manuscript.\\

\noindent \sb{\larger[1]Competing Interests}\\
\noindent The authors declare no competing interests.\\

\noindent \sb{\larger[1] Additional Movies and Figures}\\

\noindent \href{\movieOneExp}{\sb{Movie 1.}}~
\sb{Measurement of the disordered system}. Motion of particles in the experimental system described in Fig.~\ref{fig:1-pairing}A-D.\\

\noindent \href{\movieTwoSqMelt}{\sb{Movie 2.}}~
\sb{Melting of a square lattice}. Progression of the configuration and the angle-averaged structure factor $S(k)$ for the simulation described in Fig.~\ref{fig:2-square}.\\

\noindent \href{\movieThreeAvalanch}{\sb{Movie 3.}}~
\sb{Supersonic quasiparticle avalanche}. Progression of a simulation starting from an isolated pair, a described in Fig.~\ref{fig:3-avalanche}.\\

\noindent \href{\movieFourHexMelt}{\sb{Movie 4.}}~
\sb{Melting of a hexagonal lattice}. Progression of the configuration and the angle-averaged structure factor $S(k)$ for the simulation described in Fig.~\ref{fig:5-hex_melt}.\\

\noindent \href{\movieFiveFlatband}{\sb{Movie 5.}}~
\sb{Melting by flat-band modes}. Progression of a simulation starting from an isolated pair in a hexagonal lattice, a described in Fig.~\ref{fig:6}A.\\

\noindent\sb{Figure~\ref{fig:S1-comparison}}. Comparison of experiment and simulations.\\

\noindent\sb{Figure~\ref{fig:S2-spectrum}}. The spectrum computed from simulations.\\

\noindent\sb{Figure~\ref{fig:S3-gr}}. Pair correlation function.\\

\noindent\sb{Figure~\ref{fig:S4-MSDhex}}. MSD progression, time scaled by $a^{7/2}/(R^{5/2} u)$.\\

\clearpage

\setcounter{equation}{0}
\setcounter{figure}{0}
\setcounter{table}{0}
\setcounter{page}{1}
\makeatletter
\renewcommand{\theequation}{S\arabic{equation}}
\renewcommand{\thefigure}{S\arabic{figure}}
\renewcommand{\thetable}{S\arabic{table}}  
\setcounter{section}{0}
\renewcommand{\thesection}{S-\Roman{section}}

\onecolumngrid
\begin{center}
\sb{{\large Supplemental Materials:}}\\ 
\vspace{1cm}
\sb{{\Large Quasiparticles, Flat Bands and the Melting of Hydrodynamic Matter}}
\vspace{1cm}
\end{center}

\begin{figure*}[h!]
\centering
\includegraphics[width=0.7\textwidth]{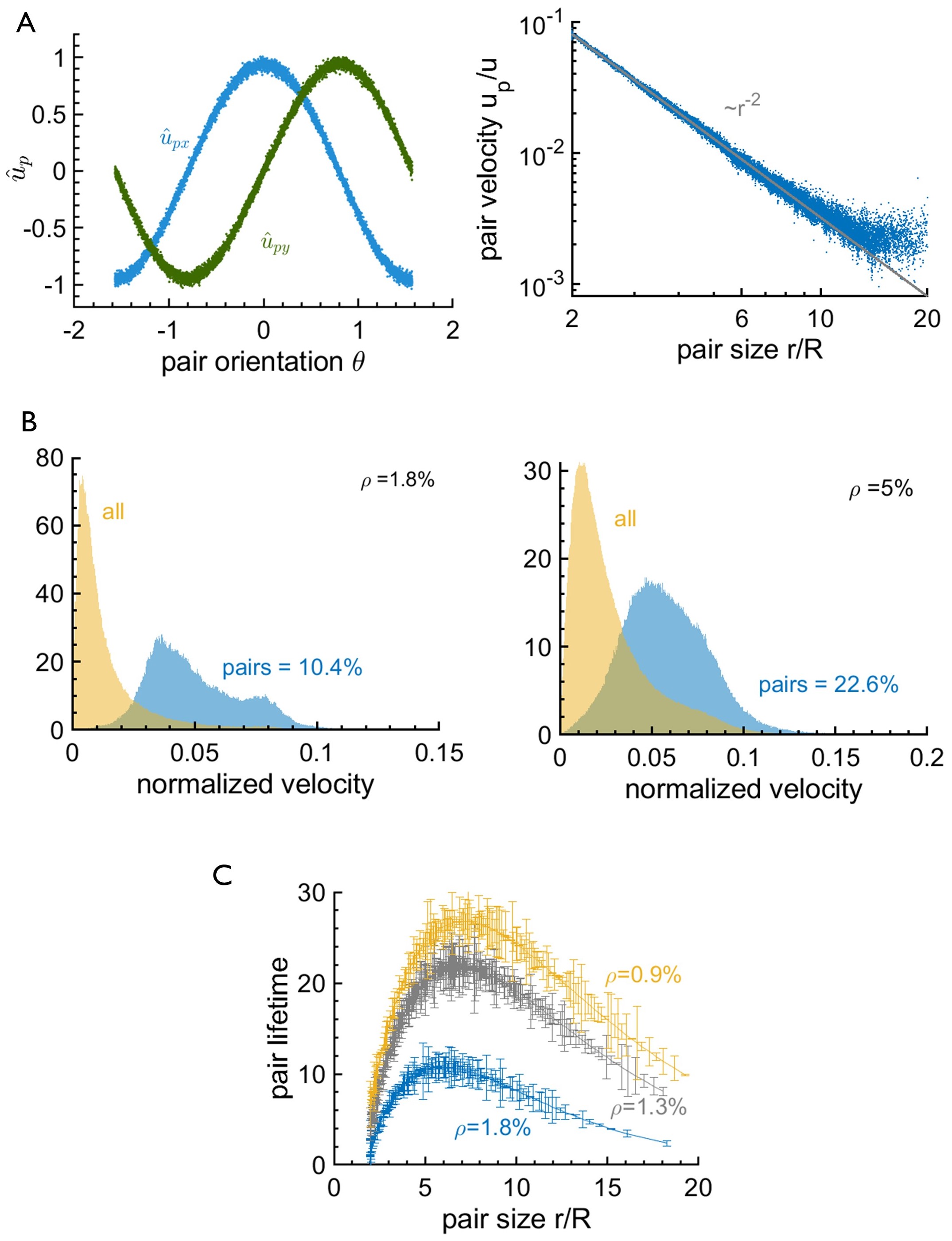}
\caption{\sb{Comparison of experiment and simulations.} 
\sb{A.}~The pair velocity $\ubp$ in simulations, showing the direction $\hat{\ub}_{\rm p} \sim (\cos{2 \theta}, \sin{2 \theta})$ (Left) and the magnitude $\abs{\ubp} \sim r^{-2}$ (Right), with a geometric factor $\alpha = 0.34$ estimated from the experiment. Solid lines are the theoretical predictions (as in the experiment, Fig.~\ref{fig:1-pairing}B).
\sb{B.}~Distribution of velocity w.r.t. center of mass (in units of $u$) of all particles (gold) and in the pairs (blue) in simulations, for areal densities $\rho = \SI{1.8}{\percent}$ (left), and $\SI{5.0}{\percent}$ (right) (as in the experiment, Fig.~\ref{fig:1-pairing}C).
\sb{C.}~Lifetime of pairs (in $R/u$ units) as a function of pair size $r/R$ for for areal densities $\rho =$ \SIlist{0.9;1.3;1.8}{\percent} in  simulations
(as in the experiment, Fig.~\ref{fig:1-pairing}D).
}
\label{fig:S1-comparison}
\end{figure*}

\begin{figure*}[ht!]
\centering
\includegraphics[width=0.7\textwidth]
{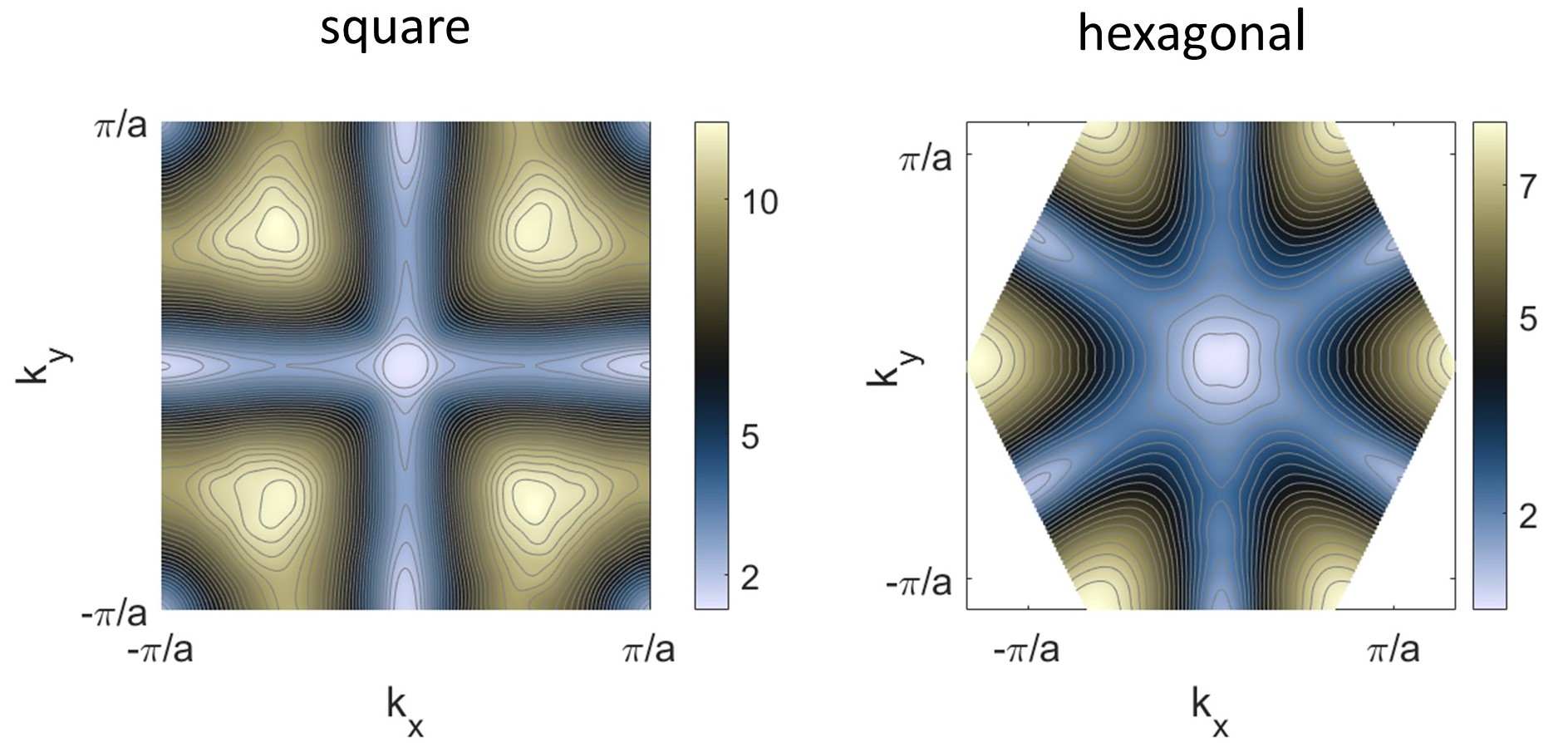}
\caption{\sb{The spectrum computed from simulations.} 
The spectrum $\omk^+ = -\omk^-$ computed in simulations of 
square (left), and hexagonal (right) lattices (Methods). Compare to theoretical spectra (compare to Fig.~\ref{fig:1-pairing}E,F and Fig.~\ref{fig:4-hex_spec}). Both lattices include $51\times 51$ particles.  
}
\label{fig:S2-spectrum}
\end{figure*}

\begin{figure*}[hb!]
\centering
\includegraphics[width=0.7\textwidth]
{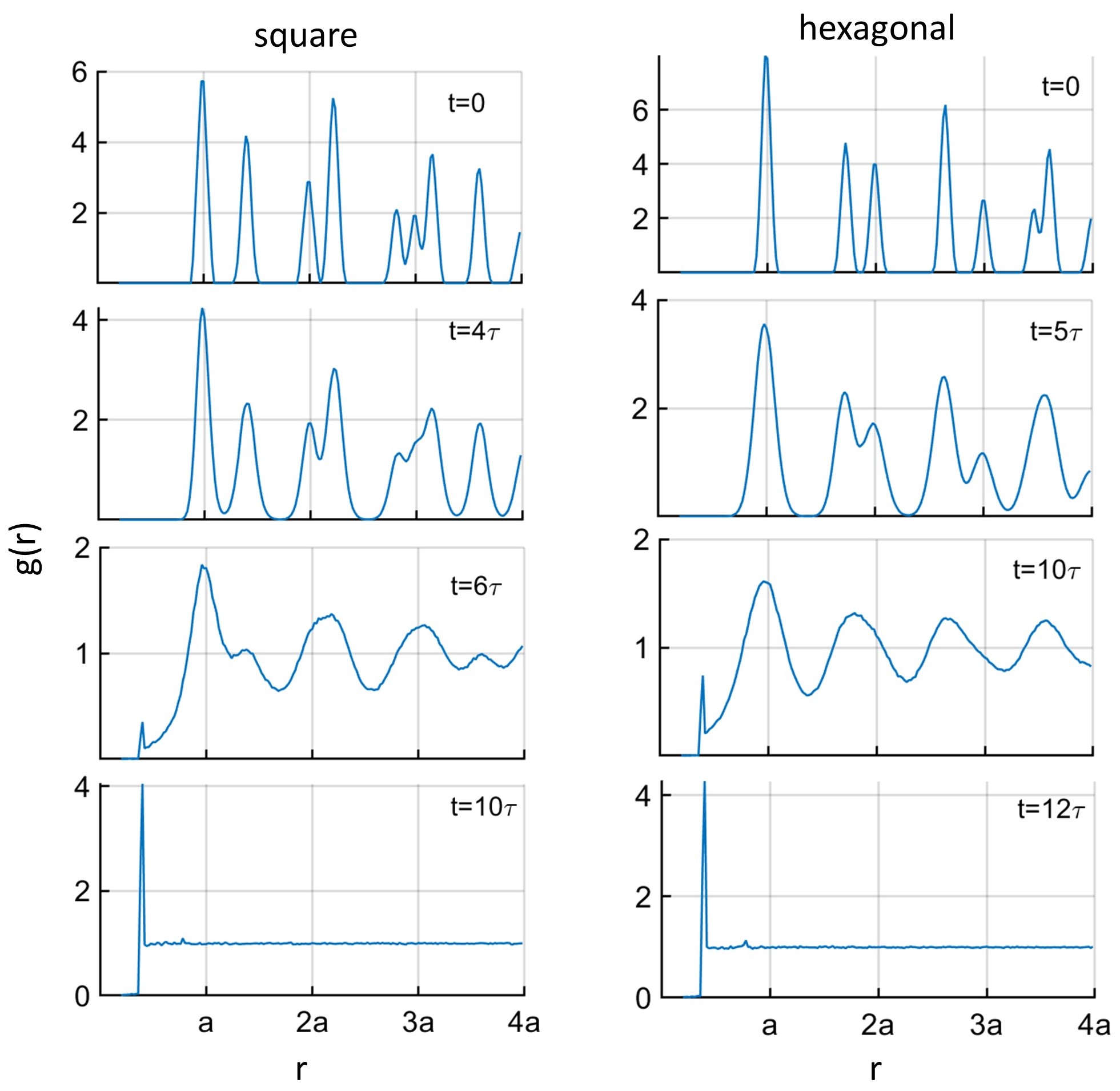}
\caption{\sb{Pair correlation function.}
The pair correlation function $g(r)$ computed in simulations of square (left) and hexagonal (right) lattices (Methods).
Times are measured in units of $\tau = a^3/(u \ell^2)$.
Both lattices include $51\times 51$ particles. 
}
\label{fig:S3-gr}
\end{figure*}

\begin{figure*}[h!]
\centering
\includegraphics[width=0.5\textwidth]{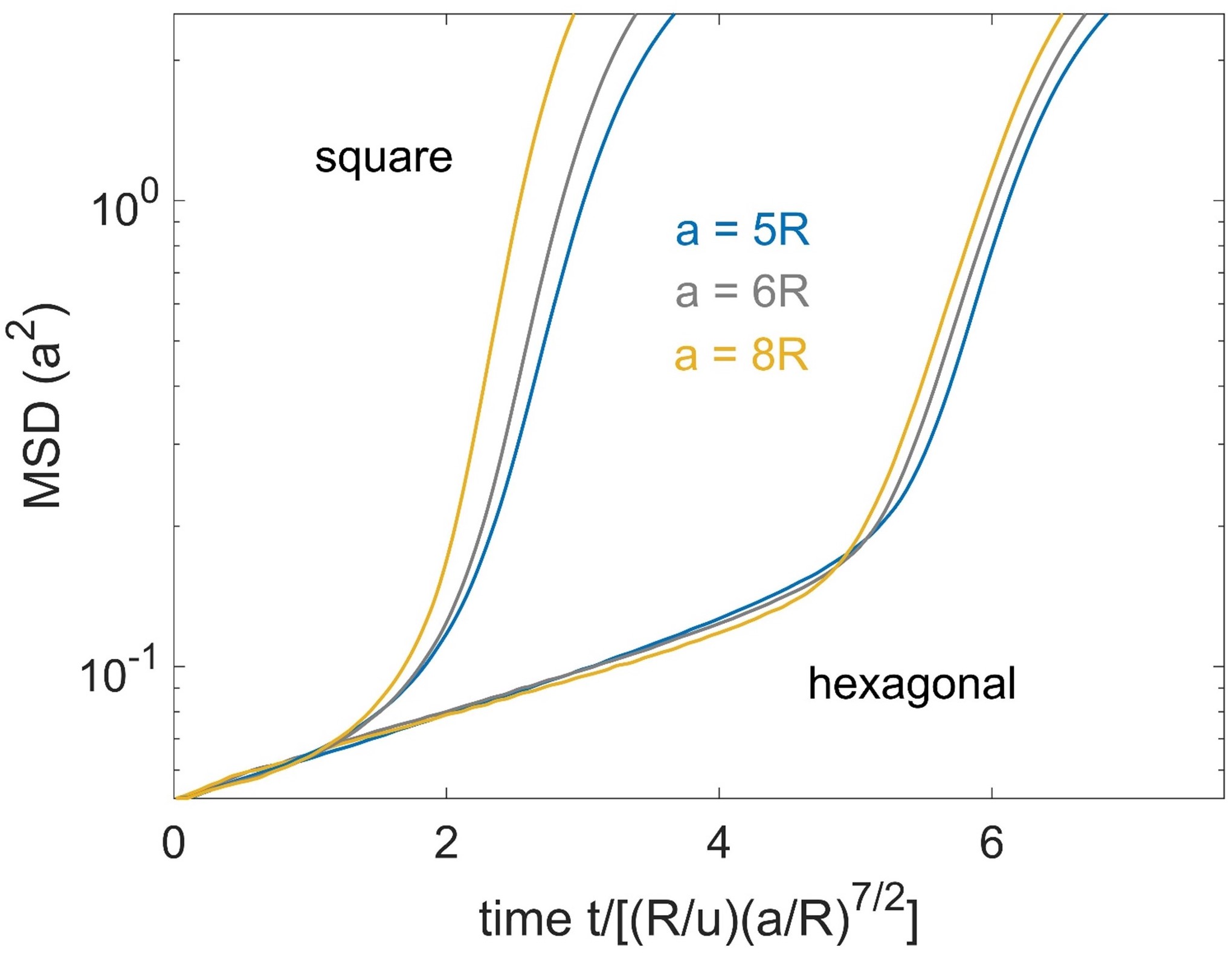}
\caption{
Progression of the mean squared deviation (MSD) in square and hexagonal crystals for $a/R = 5,6,8$, where time is normalized by $(R/u)(a/R)^{7/2} = \tau (a/R)^{1/2}$.   
}
\label{fig:S4-MSDhex}
\end{figure*}

\end{document}